\documentclass[review]{elsarticle}
 \usepackage{graphicx}
 \usepackage{amsmath,amssymb}
 \usepackage{amsmath}
\usepackage{lineno,hyperref}
\modulolinenumbers[5]

\newcommand{\bsbeq}{\begin{subequations}}
\newcommand{\esbeq}{\end{subequations}}
\newcommand{\beq}{\begin{equation}}
\newcommand{\eeq}{\end{equation}}
\newcommand{\beqa}{\begin{eqnarray}}
\newcommand{\eeqa}{\end{eqnarray}}
\newcommand{\eq}[1]{(\ref{eq:#1})}
\newcommand{\dfeq}[1]{\label{eq:#1}}

\newcommand\bbR{{\mathbb{R}}}
\newcommand\bbN{{\mathbb{N}}}
\newcommand\sgn{{\mathrm{sgn}}}

\newtheorem{lem}{Lemma}
 \newtheorem{rem}{Remark}
 \newtheorem{thm}{Theorem}
 \newtheorem{pf}{Proof}

\journal{Journal of \LaTeX\ Templates}

\begin{document}
\begin{frontmatter}
\title{Adaptive Gains to Super-Twisting Technique for Sliding Mode Design }


 \author[1,4]{Xiaogang Xiong} 
 \author[2]{Shyam Kamal}
 \author[3]{Shanhai Jin}
\address[1]{Harbin Institute of Technology, Shenzhen, P.R.China}
\address[4]{Shenzhen Key Lab of Mechanisms and Control in Aerospace, Shenzhen, P.R.China}
 \address[2]{Department of Electrical Engineering Indian Institute of Technology (BHU), Varanasi Uttar Pradesh-221005, India}
 \address[3]{School of Engineering, Yanbian University, Yanji 133002, China}








\begin{abstract}
This paper studies the super-twisting algorithm (STA) for adaptive sliding mode design. The proposed method tunes the two gains of STA on line simultaneously such that a second order
sliding mode can take place with small rectifying gains. The perturbation magnitude is obtained exactly by employing a third-order sliding mode observer in opposition to the conventional approximations by using a first order low pass filter. While driving the sliding variable to the sliding mode surface, one gain of the STA automatically converges to an adjacent area of the perturbation magnitude in finite time. The other gain is adjusted by the above gain to guarantee the robustness of the STA. This method requires only one parameter to be adjusted. The adjustment is straightforward because it just keeps increasing until it fulfills the convergence constraints. For large values of the parameter, chattering in the update law of the two gains is avoided by employing a geometry based backward Euler integration method. The usefulness is illustrated by an example of designing an equivalent control based sliding mode control (ECBC-SMC) with the proposed adaptive STA for a perturbed LTI system.
\end{abstract}


\end{frontmatter}


\section{Introduction}
\label{sec:intro}
Sliding mode control (SMC) has been recognized as one of potentially useful control schemes due to
its finite-time convergence, tracking accuracy and robustness against uncertainty \cite{Davila_2005_Oberver,Moreno_2008_lyapunov,Levant_1993_Order,Levant_2007_principle}. In practice, the main drawback of SMC is numerical chattering which could cause damages to the actuators of systems. Several solutions have been proposed to alleviate the numerical chattering, such as backward Euler methods \cite{Acary_2012_Euler,Acary_2010_Euler}, higher order sliding mode (HOSM) \cite{Levant_2007_principle,Shyam_2015_Continuous} and adaptive sliding mode designs \cite{Taleb_2014_HOSMC,Utkin_2013_ASTA}.

\par

Implicit Euler methods can totally remove the chattering and attenuate disturbances to the level of the sampling time to power level of the highest order of sliding mode, which is comparable to the conventional explicit Euler methods. However, implicit Euler methods currently are limited on only simple structures of SMC, such as the first order sliding and second order twisting controller \cite{Wang_2015_SMC,Huber_2014_Twisting}. For more complicated structure of higher sliding mode, comparing with explicit Euler methods, they needs additional special solvers to obtain chattering free solutions \cite{Acary_2012_Euler,Acary_2010_Euler}.

\par
Employing higher order sliding mode can efficiently remove the chattering \cite{Levant_2003_higher,Levant_2007_principle,Laghrouche_2007_integral}. However,
the implementation of higher order sliding mode requires
the higher order derivatives of the sliding variables and the
upper bounds of the perturbations. In practical applications, it is hard to obtain the knowledge of the bounds. Very large gain magnitudes can be selected to be greater than the actual boundaries of perturbations, satisfying the convergence requirement. However, large gain magnitudes lead to large chattering magnitudes, which is one of the reasons for the development of adaptive gains.

\par

The adaptive sliding mode is to render gains adaptive in the conventional SMC.
Since the magnitude of chattering is proportional to the gains, the chattering effect can be reduced if the gains automatically fit themselves to perturbations the SMC needs to counteract. Some adaptive laws for sliding mode control have been proposed for first order and higher order sliding mode control, e.g., \cite{Plestan_2010_ASMC,Shtessel_2012_ASTA,Taleb_2014_HOSMC}. Knowledge of perturbations is not required in such adaptive schemes. The adaptation is introduced to only one gain, i.e., the gain of highest relative order variable. However, it is worth noticing that the mechanism of second order sliding modes can have two gains corresponding to the two state variables. One of popular second order sliding mode methods is the super-twisting algorithm.
Unlike other higher order sliding model, super-twisting algorithm (STA) only needs the measurability of
the sliding variable. This characteristic make it widely used
in sliding mode control design \cite{Levant_2007_principle}, observer design
\cite{Davila_2005_Oberver} and differentiators
\cite{Levant_1998_Differentiation}.

\par
For the STA,
several adaptive laws have been developed based on the concept of
equivalent control as well as Lyapunov functions directly
\cite{Shtessel_2012_ASTA,Utkin_2013_ASTA,Utkin_2007_ASTA,Edwards_2014_Twisting,Edwards_2015_Twisting}.
The approach in \cite{Shtessel_2012_ASTA} adapts the two gains to
perturbations based on Lyapunov functions. It, however sometimes results in underestimation. Another kinds of adaptation laws in \cite{Utkin_2013_ASTA,Edwards_2014_Twisting,Edwards_2015_Twisting} based on the idea of equivalent control has a feature that asymptotically converges an adaptive gain to the minimum possible magnitude to achieve the second order sliding mode in the presence of perturbations. They, however, are based on the assumption that the equivalent control can be approximated by low-pass filtering. The values of parameters in low-pass filters give great influences to the approximation of equivalent control for systems with different noise magnitudes, actuator properties and sensor characteristics.


\par

This paper removes the usage of low-pass filters to approximate the equivalent control. A third-order sliding mode observer is employed to exactly obtain the magnitude of perturbation, which is the minimum level of the two gains in STA. Taking into account the practical advantage of seeking the minimum possible gain, similar mechanism of the adaptive SMC proposed in \cite{Utkin_2013_ASTA} has been employed. Nevertheless, here, both two gains are rendered to be adaptive by using theorems of guaranteeing robustnesses of STA in \cite{Moreno_2008_lyapunov,Moreno_2012_strict}. One gain is updated online in accordance with the adaptation of the other gain based on Lyapunov-type analysis that guarantees robustness with respect to perturbations. This paper is organized as follows: Section~\ref{sec:problem} introduces the system and the problem to be tackled. The main results are given in Section~\ref{sec:main}. An example is provided by designing a conventional equivalent control based sliding mode controller (ECB-SMC) with the proposed adaptive gains for a LTI sytem in Section~\ref{sec:exam}, and the effectiveness of the proposed method is illustrated. Finally, Section~\ref{sec:con} draws some conclusions.

\section{Problem statement}
\label{sec:problem}
The super-twisting algorithm (STA) introduced in \cite{Levant_1998_Differentiation} is one of popular second order sliding mode algorithms. It is based on the following second order system\footnote{Here, the conventional differential equations of STA are written as differential inclusion \eq{STA} with the symbol ``$\in$" because the set-valued essences of signum function \eq{sgn} will be employed for the implicit Euler integration in the subsequence. }:
\bsbeq \dfeq{STA}
\begin{align}
\dot z_1&= -\alpha |z_1|^{1/2}\sgn (z_1)+z_2 \dfeq{STA_a} \\
\dot z_2&\in -\beta \sgn(z_1)+\rho_0(t) \dfeq{STA_b}
\end{align}
\esbeq
where $\sgn(\cdot)$ is defined as a set-valued inclusion instead of single-valued function \cite{Acary_2008_book,Kikuuwe_2015_lead}:
\begin{align} \dfeq{sgn}
\mathrm{ {sgn}}(z_1):=\begin{cases}
z_1/|z_1|, &\mbox{if } z_1\neq 0,\\ [-1,1], &\mbox{if } z_1=0.
\end{cases}
\end{align}
Scalars $z_i(t) \in \bbR, i \in \{1, 2\}$ are variables with respect
to $ t \in R_+ := [0,\infty) $, and they are packed into the
state vector $z(t) = [z_1(t), z_2(t)]^T$. Positive real numbers
$\alpha$ and $\beta$ represent gains. The functions $\rho_0( t)$
denotes a perturbation satisfying
\bsbeq \dfeq{Rho}
\begin{align}
 |\rho_0( t)|&\leq L_1, \forall z \in \bbR ^2, t \in \bbR_+ \dfeq{Rho_a} \\
 |\rho_1(t)|&\leq L_2,  \forall z \in \bbR ^2, t \in \bbR_+ \dfeq{Rho_b}
\end{align}
\esbeq
where $\rho_1(t):=\dot \rho_0(t)$ and $L_1, L_2$ are non-negative finite scalars. Such an assumption of limit-size perturbation is reasonable and
similar ones can be found in \cite{Utkin_2013_ASTA,Edwards_2015_Twisting,Moreno_2012_strict}. By the solution
of a differential inclusion with discontinuous right
hand side such as \eq{STA}, it means the Filippov solution here.

\par

The objective of STA is to drive $z$ to zero in finite
time in the presences of $\rho(z, t)$ with appropriately selected the gains
$\alpha, \beta$, i.e., achieving second order sliding mode control with only the knowledge of $z_1(t)$. The purpose of this paper is to
develop an adaptive scheme allowing the gains $\alpha$ and
$\beta$ to be time-varying and updated on-line to set the gains as low as possible. The adaptation aims
at the attenuation of chattering and adjustment of parameters, which typical sliding
mode algorithms suffer from.

\section{Adaptive-gain design}
\label{sec:main}
This section proposes an approach to simultaneously adjusting
gains $\alpha$ and $\beta$ in \eq{STA} on-line to drive the
state vector $z$ to the origin precisely in finite time.
The gains are automatically increased when perturbation
$\rho_0(t)$ is large.
The adaptation also allows the gains $\alpha$ and $\beta$ to reduce
automatically if a bound of the perturbation given a priori is
too large. This section starts with an estimation mechanism of $\rho_0(t)$. Then a variable gain algorithm is proposed. At last, the new adaptive mechanism will be introduced later.

\subsection{Disturbance $\rho_0(t)$ estimation}\label{ESTIMATION}
To reduce the gains $\alpha$ and $\beta$ as much as possible with maintaining the robustness of STA, the only way is to make $\alpha$ and $\beta$ slightly greater than the necessary level such that the disturbance $\rho_0(t)$ is counteracted. This requires the exact estimation of the magnitude of $\rho_0(t)$. In the literature of estimating perturbations \cite{Utkin_2013_ASTA,Xiong_2014_Twisting,Edwards_2014_Twisting,Edwards_2015_Twisting}, a low-pass filter is usually used to approximate the equivalent control of $\rho_0(t)$, i.e., $\beta\sgn(z_1)|_{eq}=\rho_0(t)$:
\begin{align} \dfeq{low_pass}
\begin{split}
\tau\dot w +w\in\beta\sgn(z_1 )
\end{split}
\end{align}
with a constant $\tau>0$ and an initial condition $w(0)=w_0\in\bbR$. The solution $w$ can be considered as the approximation of $\rho_0(t)$, i.e., $w\approx \beta\sgn(z_1)|_{eq}=\rho_0(t)$ when $\tau$ is small enough.
The problem of such method is that the parameter $\tau$ has a very strong influence on the output $w$ and it is difficult to select the value for different systems with different sampling-time step sizes, noise magnitudes and actuator characteristics.

\par
Here, an observer is proposed to remove the usage of low-pass filter but to estimate $\rho_0(z,t)$ precisely. The observer is \cite{Shyam_2014_Twisting}
\bsbeq \dfeq{Observer}
\begin{align}
\dot {\hat z}_1&= -\alpha |z_1|^{1/2}\sgn(z_1)+\hat z_2 +k_1|e_1|^{2/3}\sgn(e_1) \dfeq{Observer_a} \\
\dot {\hat z}_2&\in -\beta \sgn(z_1)+k_2|e_1|^{1/3}\sgn(e_1)+\hat z_3   \dfeq{Observer_b} \\
\dot {\hat z}_3&\in k_3\sgn(e_1) \dfeq{Observer_c}
\end{align}
\esbeq
where $e_1:=z_1-\hat z_1$, $e_2:=z_2-\hat z_2$ and $k_1,k_2, k_3$ are appropriate positive constants. The state $\hat z_3$ is the exact estimation of $\rho_0(t)$ after a finite time $t_e>0$, i.e., $\forall t\in [t_e,+\infty), \hat z_3=\rho_0(z,t)$ and the explanation is as follows. The corresponding error dynamics of the observer \eq{Observer} is
\bsbeq \dfeq{Error}
\begin{align}
\dot e_1&= e_2-k_1|e_1|^{2/3}\sgn(e_1) \dfeq{Error_a} \\
\dot e_2&=\rho_0(z,t)-k_2|e_1|^{1/3}\sgn(e_1)-\hat z_3  \dfeq{Error_b} \\
\dot {\hat z}_3&\in k_3\sgn(e_1). \dfeq{Error_c}
\end{align}
\esbeq
By defining $e_3:=\rho_0(z,t)-\hat z_3$, equation \eq{Error} can be rewritten as
\bsbeq  \dfeq{Error_standard}
\begin{align}
\dot e_1&= -k_1 |e_1|^{2/3}\sgn(e_1)+e_2 \dfeq{Error_standard_a} \\
\dot e_2&= -k_2|e_1|^{1/3}\sgn(e_1)+e_3 \dfeq{Error_standard_b} \\
\dot e_3&\in -k_3\sgn(e_1)+\rho_1(z,t). \dfeq{Error_standard_c}
\end{align}
\esbeq
According to \cite{Levant_2003_higher}, one can use the homogeneity property of \eq{Error_standard} to select the values of $k_1$, $k_2$ and $k_3$ such that
the three share one common parameter $L>0$, and correspondingly the error $e:=[e_1,e_2,e_3]^T$ in \eq{Error_standard} converges to zero in finite time $t\in [t_e,+\infty)$ if $L>L_2\geq |\rho_1(z,t)|$. The convergence time $t_e$ is an inverse function of $L$ and proportional function of $e(0)$. For example, in \cite{Levant_2003_higher}, the three parameters are chosen as
\begin{align} \dfeq{gains}
k_1=3L^{1/3}, k_2=1.5\sqrt{3}L^{2/3}, k_3=1.1L.
\end{align}
Therefore, for $t\in [t_e,+\infty)$, $\hat z_3$ is viewed as the equivalence control of the perturbation $\rho_0(z,t)$.

\subsection{Selection of $\alpha$ and $\beta$}
In this section, a variable gain-selection algorithm for STA \eq{STA} is introduced to give additional choices for updating $\alpha$ and $\beta$ without damaging the robustness.

\begin{lem}\cite{Moreno_2012_strict}\label{lem}
Suppose the perturbation term $\rho_0(z,t)$ in the STA \eq{STA} globally bounded by \eq{Rho}. Then for every positive $L_1>0$, there exists a pair of gains $\alpha$ and $\beta$ such that $z=0$ is a robustly and globally
finite-time stable equilibrium point. Moreover, there exists a symmetric
and positive definite matrix $P=[p_{ij}]\in \bbR^{2\times 2}$ such that $V(z)=\zeta^T P \zeta$ with $\zeta:=[|z_1|^{1/2}\sgn(z_1), z_2]^T$ is a quadratic, strict and robust Lyapunov function for the perturbed system \eq{STA},
satisfying
\begin{align}
\dot V
\leq -\frac{1}{|z_1|^{1/2}}\zeta^T Q_R\zeta
\dfeq{comparez_original}
\end{align}
almost everywhere, for some symmetric and positive definite matrix $Q_R:=[q_{Rij}]\in \bbR^{2\times 2}$. Furthermore, a trajectory starting at $z_0$ will converge to the origin in a finite time smaller than $t_z(z_0)$:
\begin{align} \dfeq{Moreno_time}
t_z(z_0)=\frac{2}{\gamma}V^{1/2}(z_0), \quad \gamma=\frac{\omega^{1/2}_{\min}\{P\}\omega^{1/2}_{\min}\{Q_R\}}{\omega_{\max}\{P\}}
\end{align}
where $\omega_{\min}\{P\}$ and $\omega_{\max}\{Q_R\}$ represent the minimum and maximum eigenvalue of $P$ and $Q_R$, respectively.
\end{lem}
The proof of Lemma~\ref{lem} in \cite{Moreno_2012_strict} derives the following algorithm for the selection rule of $\alpha$, $\beta$, $p_{ij}$ and $q_{Rij}$:
\begin{description}
  \item[(i)] Choose positive constants $(\lambda,h)$ so that $0<\lambda<1$ and $h>1$.
  \item[(ii)] Find positive constants $(\theta_1,\theta_2)$ satisfying the inequality
\beqa \dfeq{parameter}
&\theta_1-\dfrac{2\theta_2}{h}>\dfrac{1}{4}\left(1+\theta_1\right)^2
  -\left(1+\theta_1\right)\theta_2\lambda+\theta_2^2.\,
\eeqa
The inequality \eq{parameter} represents the interior of an
ellipsoid on the $(\theta_1,\theta_2)$-plane parameterized by
$h$ and $\lambda$.
Indeed, it can be transformed into the following standardized formulation:
\begin{align} \dfeq{ellipse}
&
\left(\dfrac{1+\theta_1}{2}\right)^2-2\left(\dfrac{1+\theta_1}{2}\right)\theta_2 \lambda+\theta_2^2
\nonumber\\
& \hspace{10ex}-2\left(\dfrac{1+\theta_1}{2}\right)
+\dfrac{2}{h}\theta_2+1<0.
\end{align}
This ellipsoid can be utilized to pick
$\theta_1$ and $\theta_2$ satisfying \eq{parameter} as
proposed in \cite{Moreno_2012_strict}.
The center of the ellipsoid \eq{ellipse} is computed as
$((1+\bar \theta_1)/2,\bar \theta_2)$, where
\begin{align}\dfeq{center}
\bar \theta_1&=\dfrac{h-2\lambda+h\lambda^2}{h(1-\lambda^2)},
\quad
\bar \theta_2=\dfrac{\lambda h-1}{h(1-\lambda^2)} .
\end{align}
If $\theta_1$ and $\theta_2$ are selected as $\bar \theta_1$ and $\bar \theta_2$, respectively, the pair
obviously satisfies \eq{parameter} or \eq{ellipse}.
Properties $\theta_1>0$ and $\theta_2>0$ are achieved
if the center of the ellipsoid is in the first quadrant of
the $(\theta_1,\theta_2)$-plane.
In fact, the positiveness of $\bar \theta_1$ and $\bar \theta_2$
can be guaranteed by choosing $h\lambda>1$.
  \item[(iii)] Given such values of $(\lambda,h)$ and $(\theta_1,\theta_2)$, the gains
  \begin{align} \dfeq{gain_moreno}
\beta=\frac{1+\lambda}{1-\lambda}{L_1}, \quad
\alpha= \theta_1\sqrt{\frac{2h}{ (1-\lambda)\theta_2}}{L_1}^{1/2}
\end{align}
assure the robust, finite-time stability of the origin of the STA \eq{STA}.
\end{description}

\par
After obtaining the constants $(h,\lambda)$, $(\theta_1, \theta_2)$ and gains $(\alpha, \beta)$, the value of $t_z(z_0)$ can be calculated by using \eq{Moreno_time} with the matrices $P$ and $Q_R$ given by:
\begin{align} \dfeq{symbols_moreno}
& p_{11}= 1,\, p_{22}=\frac{(1-\lambda)\theta_2}{2L_1}, \, p_{12}:= -\sqrt{\frac{p_{22}}{h}},\nonumber \\
& q_{R11}=\alpha+2p_{12}(\beta+L_1)+2L_1(1-\alpha p_{12})\frac{p_{22}}{p_{12}} \nonumber \\
 & q_{R12}=-\frac{1}{2}(1-\alpha p_{12})+(\beta+L_1)p_{22},\, q_{R22}=-p_{12}.
\end{align}
\par
It should be noted that for a constant value of $h>1$, the size of ellipse \eq{ellipse} is solely determined by the value of $\lambda$. If the inequality \eq{ellipse} is satisfied by a given value $\lambda=\lambda_m<1$, then, for a function $\lambda(t)\geq \lambda_m$ replacing $\lambda_m$ in \eq{parameter} and \eq{ellipse}, the inequality \eq{parameter} and \eq{ellipse} is still satisfied. The new center related point $(\bar\theta_1(t),\bar \theta_2(t))$ defined by $h$ and $\lambda(t)$ as in \eq{center} is always located within the new size-variable ellipse \eq{ellipse}. Based on this observation, a modified version of Lemma~\ref{lem} is provided here to render $\alpha$ and $\beta$ adaptive without loss the robustness of the perturbed STA \eq{STA}. Here, $\rho_0(z, t)$ in \eq{Rho} now is assumed to be exactly observable.
 The new algorithm is as follow:

\begin{description}
  \item[(i)] Choose positive constants $(\eta,h, p_{22})$ so that
\begin{align} \dfeq{new_inequality}
& 0<\eta< 1, \quad  h>1, \quad  p>0
\end{align}
  \item[(ii)] The gain $\beta(t)$ is assigned here as
\begin{align}\dfeq{beta_new}
&\beta(t)\!=\max\left(\beta_m,\,\frac{|\rho_0(z,t)|}{\eta}\right)
\end{align}
with a positive constant $\beta_m>0$ based on the assumption of the observability of $\rho_0(z,t)$.
Then, calculate the positive variable $\lambda(t)$ and the variable center point $(\bar \theta_1(t),\bar\theta_2(t))$：
\begin{align}\dfeq{center_new}
\bar \theta_2(t)&=\beta(t)p \nonumber\\
\lambda(t)\!&=\! -\frac{1}{2\bar\theta_2(t)}+\frac{\sqrt{h^2+4\bar\theta_2(t) h+4\bar\theta(t)^2h^2}}{2\bar\theta_2(t)h}\nonumber \\
\bar \theta_1(t)&=\dfrac{h-2\lambda(t)+h\lambda(t)^2}{h(1-\lambda(t)^2)}.
\end{align}
The positiveness of $\bar \theta_1(t)$ and $\bar \theta_2(t)$
can be guaranteed by choosing $h>1$. Actually, the variable point $((1+\bar \theta_1(t))/2,\bar \theta_2(t))$ is the center of the following size-variable ellipsoid
\beqa \dfeq{parameter_new}
&\theta_1(t)-\dfrac{2\theta_2(t)}{h}>\dfrac{1}{4}\left(1+\theta_1(t)\right)^2+\theta_2(t)^2
   \nonumber \\
&\hspace{25ex}-\left(1+\theta_1(t)\right)\theta_2(t)\lambda(t).\,
\eeqa
  \item[(iii)] Given such values of $(\eta, h, p)$ and functions $\lambda(t)$, $(\bar \theta_1(t),\bar \theta_2(t))$, the gain $\alpha(t)$
  \begin{align} \dfeq{gain_moreno_new}
  \alpha(t)= \bar\theta_1(t)\sqrt{\frac{h}{p}}.
\end{align}
 assures the robust, finite-time stability of the origin of the STA \eq{STA}. Obviously, $p$ is used to adjust the value of $\alpha(t)$.
\end{description}
\begin{thm}: \label{thm1}
Consider \eq{STA} satisfying \eq{Rho} with gains rendered to be variable, i.e., $\alpha(t)>0$, $\beta(t)>0$. Given the perturbation $\rho_0(z,t)$ observable, if $\beta(t)$ and $\alpha(t)$ selected as in\eq{center_new} and \eq{gain_moreno_new}
with $\{\eta,\lambda(t), h, \bar \theta_1(t),\bar \theta_2(t)\}$ satisfying \eq{new_inequality}-\eq{center_new}, then $z=0$ is globally finite time stable and
there exists $t_z(z_0)\in\bbR_+$ such that $z=0$ is a robustly and globally finite-time stable equilibrium point.
Moreover, the function $V(z)=\zeta^T P \zeta$ with $\zeta:=[|z_1|^{1/2}\sgn(z_1), z_2]^T$ and $P=[p_{ij}]$ defined by \eq{symbols_moreno_new} is a quadratic, strict and robust Lyapunov function for the perturbed system \eq{STA} satisfying \eq{Rho} with observable $\rho_0(z,t)$ and variable gains $\alpha(t)$ and $\beta(t)$. It satisfies the inequality:
\begin{align}
\dot V
\leq -\frac{1}{|z_1|^{1/2}}\zeta^T Q_R(z,t)\zeta
\dfeq{comparez_new}
\end{align}
almost everywhere, for $Q_R(z,t):=[q_{Rij}]$ given in \eq{symbols_moreno_new}. Furthermore, a trajectory starting at $z_0$ will converge to the origin in a finite time smaller than $t_z(z_0)$:
\begin{align} \dfeq{SMTW}
&z(t)=0, \quad \forall t\in [t_z(z_0),\infty ) \nonumber \\
\end{align}
where $t_z(z_0)$ is defined according to \eq{Moreno_time}.
\end{thm}
\begin{pf}
The vector $\zeta$ means that $z=0$ if and only if $\zeta=0$, because it is a bijective map between $z\in \bbR^2$ and $\zeta\in \bbR^2$.
As done in \cite{Moreno_2008_lyapunov,Moreno_2012_strict}, define $\rho_z(z,t)$ and $A(z,t)$ by
\begin{align}
&
\rho_0(z,t)=\rho_z (z,t)\sgn(z_1)
\dfeq{nuzt}
\\
&
A(z,t)=\begin{bmatrix}
       \ -\alpha(t)            & 1 \          \\[0.3em]
       \ -2 \beta(t)+2\rho_z(z,t)  & 0 \
     \end{bmatrix} .
\dfeq{Azt}
\end{align}
Property \eq{Rho} results in
\begin{align} \dfeq{nu_condition}
\rho_z(z,t)=\rho_0(z,t)\sgn(z_1), \forall z\in \bbR^2\setminus \{z_1=0\},t\in \bbR_+.
\end{align}
Let $I$ denote the identity matrix of appropriate dimension. Consider a constant matrix $P$ written as
\begin{align}
\dfeq{P}
P=\begin{bmatrix}
p_{11} & p_{12} \\
p_{12} & p_{22}
\end{bmatrix}
\end{align}
and let $Q(z,t)$ be defined with
\begin{align}
-Q(z,t)=PA(z,t)+A(z,t)^TP .
\dfeq{LinLyapQ}
\end{align}
Then the matrix $Q(z,t)$ is computed as
\begin{align} \dfeq{Q_20}
 &Q(z,t)\!=\!\begin{bmatrix}
      2\alpha p_{11}+4p_{12}(\beta(t)\! -\!\rho_z(z,t)) & \star          \\[0.3em]
       \alpha p_{12}+2p_{22}(\beta(t)\! -\!\rho_z(z,t) )\!\!-p_{11}& \!-\!2p_{12}           \\[0.3em]
     \end{bmatrix}.
\end{align}
Here, the symbol ``$\star$" denotes a symmetric component.
The matrices $P$ and $Q$ satisfy
\begin{align}
& P>0
\dfeq{PQP}
\\
& Q(z,t)\geq \omega(t) I
, \quad  \forall (z,t)\in\bbR^2\setminus\{z_1=0\}\times\bbR_+
\dfeq{PQQ}
\end{align}
for some $\omega(t) >0$ if the following inequalities are satisfied
uniformly in $(z,t)\in\bbR^2\setminus\{z_1=0\}\times\bbR_+$:
\begin{align}
&
p_{11}=1, \quad p_{22}>p_{12}^2,\quad p_{12}<0
\dfeq{conditionsa}
\\
&
0>\!p_{11}\alpha  p_{12}+\!\frac{1}{4}(p_{11}\!-\!\alpha p_{12})^2
+2p_{12}^2(\beta(t)\! -\!\rho_z(z,t))-
\nonumber \\
&(p_{11}\!-\!\alpha p_{12})p_{22}(\beta(t)\! -\rho_z(z,t))+(\beta(t) \!-\!\rho_z(z,t))^2p_{22}^2.
\dfeq{conditionsc}
\end{align}
The inequalities \eq{PQP} and \eq{PQQ} are satisfied for some $\omega(t) >0$ if
\eq{conditionsa}-\eq{conditionsc} are met.
Due to \eq{beta_new} and \eq{nu_condition},
%
the inequality \eq{conditionsc} is satisfied uniformly in $(z,t)$ if
\begin{align} \dfeq{condition_1}
& 0>\frac{1}{4}(1-\alpha(t) p_{12})^2
+2p_{12}^2(\beta(t)-\rho_z(z,t))\nonumber\\
&\hspace{0ex}+
\alpha(t) p_{12}-(1- p_{12}\alpha(t)) (\beta(t)-\rho_z(z,t))p_{22}\nonumber\\
&+p_{22}^2(\beta(t)-\rho_z(z,t))^2.
\end{align}
Defining
\begin{align} \dfeq{symbols}
 &h:=\frac{p_{22}}{p_{12}^2},\quad \theta_1(t):= -\alpha(t) p_{12}, \, \nonumber\\
&\theta_2(t):= (\beta(t)-\rho_z(z,t))p_{22}
\end{align}
leads to
\begin{align}
&\theta_1(t)-2\frac{  \theta_2(t)}{h}>\frac{1}{4}(1+ \theta_1(t))^2-\nonumber \\
&\hspace{20ex}(1+ \theta_1(t))\theta_2(t)+\theta_2(t)^2. \dfeq{new_condition}
\end{align}
Obviously, \eq{new_condition} is satisfied if
\begin{align} \dfeq{new_condition_last}
& \theta_1(t)-2\frac{  \theta_2(t)}{h}>\frac{1}{4}(1+ \theta_1(t))^2-(1+ \theta_1(t)) \theta_2(t)\lambda(t)\nonumber \\
&\hspace{10ex}+\theta_2(t)^2, \quad 0< \lambda(t)<1
\end{align}
is met. The inequality \eq{new_condition_last} is a size-variable ellipse, of which the size is determined by $h$ and $\lambda(t)$. Similar to the algorithm of Lemma~\ref{lem}, the mission is to find suitable functions $\theta_1(t)>0$ and $\theta_2(t)>0$ that satisfy \eq{new_condition_last}. Select the center point $(\bar \theta_1(t),\bar \theta_2(t))$, which can be obtained by using the definition \eq{center} with the constant $\lambda$ replaced by the variable $\lambda(t)$, as the point $(\theta_1(t),\theta_2(t))$ in \eq{new_condition_last}. Let
\begin{align} \dfeq{new_h_condition}
&\bar \theta_2(t)=\frac{h\lambda(t)-1}{h(1-\lambda^2(t))}:=\beta(t)p_{22},
\end{align}
which means that the vertical coordinate $\bar \theta_2(t)$ of the center $((1+\bar \theta_1(t))/2,\bar \theta_2(t))$ of the ellipse \eq{new_condition_last} is forced to change according to the changes of $\beta(t)$. The equivalence \eq{new_h_condition} leads to the expression of $\lambda(t)$ in \eq{center_new}, as the input of $\bar \theta_1(t)$ in \eq{center_new}. As a result, the center point $((1+\bar \theta_1(t))/2,\bar \theta_2(t))$ of the ellipse \eq{new_condition_last} changes according to the changes of $\beta(t)$. The equivalence $\bar \theta_1(t)=-p_{12}\alpha(t)$, $h=p_{22}/p^2_{12}$ and \eq{new_h_condition} lead to the expression \eq{gain_moreno_new}. The positiveness of $(\bar \theta_1(t),\bar \theta_2(t))$ requires $h\lambda(t)>1$, which is always satisfied if $h>1$.
\par

It should be noted that the size of the ellipse \eq{new_condition_last} is consistent with the value of $\beta(t)$ and $\alpha(t)$. For a given constant $h$, the size of the ellipse \eq{new_condition_last} is determined by $\lambda(t)$, which is forced to be as in \eq{new_h_condition}. Then, the change in size of the ellipse \eq{new_condition_last} is driven by the magnitude of $\beta(t)$. The size of ellipse \eq{new_condition_last} increases as $\beta(t)$ increases, making
the inequality \eq{new_condition_last} always satisfied. If $\beta(t)$ changes according to the magnitude of $\rho_0(z,t)$, one can note that the size of the ellipse \eq{new_condition_last} is finally driven by the magnitude of $\rho_0(z,t)$.

\par
Choose $V(\zeta)\!=\!\zeta^T P \zeta$ with $P$ defined in \eq{P}, which is a constant
positive definite symmetric matrix with elements $p_{11}$, $p_{12} $ and $p_{22}$ given by \eq{conditionsa} and \eq{symbols}. By using the facts $\dfrac{d}{dt}|z_1|^{1/2}\sgn(z_1)
\!=\!\dfrac{1}{2|z_1|^{1/2}}\dfrac{d}{dt}{z}_1, \,z_1\!\neq \!0$,
$\dot \zeta\!=\!\dfrac{1}{2|z_1|^{1/2}}A(z,t)\zeta$ and \eq{LinLyapQ}, the time derivative $\dot V$ along the solutions to
\eq{STA} and \eq{nuzt} is computed as: $z_1 \neq 0 $,
\begin{align}
\dot V=
-\frac{1}{|z_1|^{1/2}}\zeta^T Q(z,t)\zeta
\leq -\frac{\omega(t) }{2|z_1|^{1/2}}\zeta^T\zeta ,
\dfeq{compare0}
\end{align}
where $\omega(t)>0$ is smallest eigenvalue of the matrix $Q_R(z,t)>0 $ given by
\begin{align}
&
Q_R(z,t)\!:=\!\begin{bmatrix}
      2\alpha(t)  +4\bar \theta_2(t)\dfrac{p_{12}}{p_{22}}+\xi(t)  & \star          \\[0.3em]
      2\bar \theta_2(t)\!-(1+\bar \theta_1(t))& \,-2p_{12}           \\[0.3em]
     \end{bmatrix}
\dfeq{QRdef}
\\
&
\xi(t) =2\left(1+\bar \theta_1(t)\right)\bar \theta_2(t)\frac{1-\lambda(t)}{p_{12}} .
\nonumber
\end{align}
The matrix $Q_R(z,t)$ is constructed by using the inequality \eq{new_condition_last} with $(\theta_1(t),\theta_2(t))$ replaced by $(\bar\theta_1(t),\bar\theta_2(t))$. It is a positive definite matrix if
\eq{new_condition_last} is satisfied and
\begin{align}
2\alpha(t)  +4\bar \theta_2(t)\dfrac{p_{12}}{p_{22}}+2\left(1+\bar \theta_1(t)\right)\bar \theta_2(t)\frac{1-\lambda(t)}{p_{12}}>0,
\end{align}
which is automatically satisfied if \eq{new_condition_last} holds.
Note that the function $V(\zeta(t))$ is proved to be
absolutely continuous in $t$ \cite{Moreno_2012_strict}, so that
$V(\zeta(t))$ is strictly decreasing in $t$ if and only if
$\dot V$ is negative definite almost everywhere. Then the time derivative $\dot V$ along the solutions to
\eq{STA} is obtained as \eq{compare0}.
From \eq{compare0}, one arrive at \eq{SMTW} with
$t_z={4\lambda_{P}^{1/2}\sqrt{V(\zeta(0))}}/{\omega(t)}$ where $\lambda_{P}$ is the smallest eigenvalue of $P$ with elements $p_{11}$, $p_{12} $ and $p_{22}$ are give as follow:
\begin{align} \dfeq{symbols_moreno_new}
 &p_{11}= 1,\,p_{22}=p,\, p_{12}= -\sqrt{\frac{p_{22}}{h}}.
\end{align}
\qed

\begin{figure*}
\begin{center}
\includegraphics[width=\textwidth]{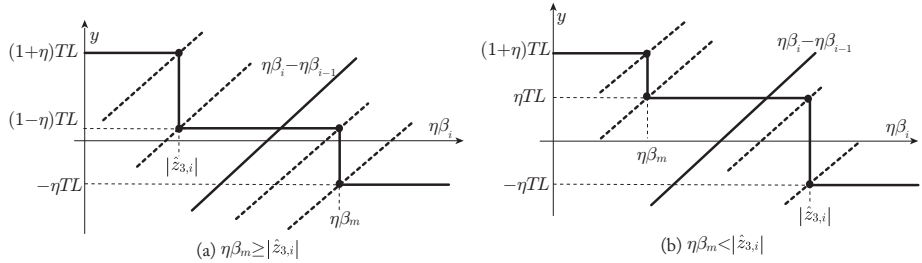}    
\caption{Illustration of the enumeration method for \eq{Discrete_form}.}  
\label{enumeration}                                 
\end{center}                                 
\end{figure*}

\end{pf}

Another selection law for $\alpha(t)$ and $\beta(t)$ can be found in \cite{Davila_2005_Oberver}, which was proven by employing the method of majorant curve. The difference is that here, instead of showing an inequality condition and constant gains, a selecting law for variable gains $\alpha(t)$ and $\beta(t)$ is given.
\par
Theorem~\ref{thm1} is a modified version of selecting $\alpha$ and $\beta$ for the STA \eq{STA} in \cite{Moreno_2012_strict}. The new added parameter $0<\eta<1$ determines how close $\beta$ is to the perturbation magnitude limitation $|\rho_0(z,t)|$ and it should be as large as possible to reduce the value of $\beta$ as much as possible, making the chattering magnitude greatly attenuated. Comparing to the original one, with the same value of $\beta$, the values of $\alpha$ selected according to Theorem~\ref{thm1} can be much smaller, reducing the chattering magnitude further.


\subsection{Update law for $\alpha$ and $\beta$}
This section introduces the approach to simultaneously adjust $\alpha$ and $\beta$ according to the perturbation $\rho_0(z,t)$, driving the state vector $z$ to the origin precisely in finite time.
Let the following adaptive mechanism be introduced to \eq{STA}:
\bsbeq \dfeq{law}
\begin{align}
 \dot \beta(t) &\in -L\sgn(\eta\beta(t)-|\hat z_3|)-\frac{L}{\eta}\mathrm{H}(\beta(t)-\beta_m)\dfeq{lawc}\\
\alpha(t)&=\bar\theta_1(t)\sqrt{\frac{h}{p}}, \quad \beta(0)=\beta_0\geq \beta_m \dfeq{lawa}
\end{align}
\esbeq
where $\alpha (t)$ and $\beta(t)$ are scalar variables, $\beta_m>0$ is a constant defined as the minimum gain magnitude of $\beta(t)$ to prevent the loss of robustness, positive constants $h$ and $p$ satisfy \eq{new_inequality}. The function $\bar \theta_1(t)$ is calculated with \eq{center_new} and $\hat z_3$ is obtained from \eq{Observer}.
The map $\mathrm H: \bbR\to\bbR$ is an inclusion modified from the Heaviside step function:
\begin{align} \dfeq{H}
\mathrm{H} (z):= \begin{cases}
0,\! &\mbox{if } z>0,\\ [-1,0],\! &\mbox{if } z=0, \\ -1,\! &\mbox{if } z<0.
\end{cases}
\end{align}

\begin{thm} \label{thm2}
Consider \eq{STA} with \eq{Observer}, \eq{gains}, and \eq{law}.
Assume that $\rho_0(z,t)$ is differentiable and satisfying \eq{Rho} with
a given variable $L_2(t)> 0$. Let
\begin{align} \dfeq{Gains}
 L> \frac{L_2(t)}{\eta},
\end{align}
then, the solutions $z(t)$, $\hat z(t)$ and $\beta(t)$ are bounded for all
$(z_0,\hat z_0,\beta_0)\in\bbR^2\times\bbR^3\times[\beta_m,+\infty )$, and the following three statements hold true:
\begin{enumerate}
    \item  There exists $t_e\in\bbR_+$ such that
$\forall t\!\in\! [t_e,\infty)$, $\hat z_3=\rho_0(z,t)$.
    \item There also exists $t_\delta\in[t_e, \infty)$ such that
\begin{align}
&\dfeq{Utkin_kTW}
\beta(t)=\max\left(\dfrac{|\rho(z,t)|}{\eta},\beta_m \right).
\end{align}
\item Furthermore, there exists $t_z\in [t_\delta, +\infty)$
\begin{align}
z(t)=0, \quad \forall t\in [t_z,\infty)
\dfeq{Convergent}
\end{align}
holds for each $(z_0,\hat z_0,\beta_0)\!\in\!\bbR^2\!\times\!
\bbR^3\!\times\![\beta_m,+\infty )$.
  \end{enumerate}
\end{thm}
%
%


\begin{pf}
First, equation \eq{Observer}\eq{Error}\eq{Error_standard} ensure
$|\hat z_3(t)|<\infty$ for all $t\in\bbR_+$ due to the finite perturbation  $|\rho_0(z,t)|\leq L_1(t)$, $|\rho_1(z,t)|\leq L_2(t)$ and conditions \eq{Gains}. The differential inclusions in \eq{law} also guarantees
$|\beta(t)|<\infty$ and $|\alpha(t)|<\infty$
for all $t\in\bbR_+$ due to $|\hat z_3(t)|<\infty$.
Moreover, the differential equations \eq{law} result in $\beta_m\leq \beta(t)<+\infty $
for all $t\in\bbR_+$ and $\beta(0)=\beta_0\geq \beta_m$.
\par
To see this, first consider $|\hat z_3|> \eta\beta_m$.
In the case of $\eta \beta>|\hat z_3|$,
$\sgn(\eta\beta-|\hat z_3|)=1$ and $\mathrm{H}(\beta-\beta_m)=0$ gives
$\dot{\beta}=-L$,
due to \eq{lawc}. Then, $\eta\beta$ approaches to $|\hat z_3|$ in finite time due the negative derivative.
In the case of $\eta \beta<|\hat z_3|$,
the definition \eq{lawc} gives $\sgn(\eta \beta-|\hat z_3|)=-1$ and $\mathrm{H}(\beta-\beta_m)=0$ leading to
$\dot{\beta}=L$. Due to $|\hat z_3|<+\infty$, $\dot{\beta}=L$ makes $\eta \beta>|\hat z_3|$ satisfied in finite time.
 In the case of $\eta\beta=|\hat z_3|$, $\sgn(\eta \beta-|\hat z_3|)\in [-1,1]$ and $\mathrm{H}(\beta-\beta_m)=0$ may lead to the decrease of $\beta$ due to the possibility of $\dot \beta<0$. However, as soon as $\beta<\beta_m$ is satisfied, one has $\sgn(\eta \beta-|\hat z_3|)\in [-1,1]$, $\mathrm{H}(\beta-\beta_m)=-1$, and $\dot \beta \in [-1,1]L+L/\eta>0 $, which leads to $\beta\geq \beta_m$ in a short time.
Therefore, in case of $|\hat z_3|>\eta\beta_m$, $\beta$ is kept in the set $\beta_m\leq \beta(t)\leq +\infty$
for all $t\in\bbR_+$.

\par
Now consider the case of $|\hat{z}_3|< \eta\beta_m$. In the case of $\beta>\beta_m$,
$\sgn(\eta\beta-|\hat z_3|)=1$ and $\mathrm{H}(\beta-\beta_m)=0$ gives
$\dot{\beta}=-L$,
due to \eq{lawc}. Then, $\beta$ approaches to $\beta_m$ in finite time due to the negative derivative. In the case of $|\hat z_3|<\eta\beta<\eta\beta_m$, $\sgn(\eta\beta-|\hat z_3|)=1$ and $\mathrm{H}(\beta-\beta_m)=-1$ gives
$\dot{\beta}=-L+L/\eta>0$, making $\beta$ to approach to $\beta_m$ in finite time due to the positive derivative. In the case of $\eta\beta<|\hat z_3|<\eta\beta_m$, $\sgn(\eta\beta-|\hat z_3|)=-1$ and $\mathrm{H}(\beta-\beta_m)=-1$ gives
$\dot{\beta}=L+L/\eta$. Therefore, in a finite time, $\beta\geq \beta_m$ is achieved.

\par
For the case of $|\hat z_3|=\eta\beta_m$, the analysis is the same as the above.

\par
Next, consider $\beta=\beta_m$.
In the case of $\eta \beta >|\hat z_3|$,
the definition \eq{lawc} gives $\sgn(\eta \beta-|\hat{z}_3|)=1$ and $\mathrm{H}(\beta-\beta_m)\in [-1,0]$, leading to $\dot{\beta}\in -L-[-1,0]L/\eta$. This means that $\beta$ may decrease due to the possibility of $\dot \beta<0$. However, as soon as $\beta$ decreases, $\beta<\beta_m$ achieves and this results in $\mathrm{H}(\beta-\beta_m)=-1$ and $\dot \beta=-L+L/\eta$, according to \eq{lawc}. This implies that $\beta$ stop decreasing because of hitting the bottom value $\beta_m$. It also may increase due to the possibility of $\dot \beta>0$. As soon as $\beta$ increase to $\beta>\beta_m$, one has $\dot{\beta}= -L$. Therefore, $\beta=\beta_m$ is kept for $\eta \beta >|\hat z_3|$.
In the case of $\eta \beta <|\hat z_3|$,
the definition \eq{lawc} gives $\sgn(\eta \beta-\hat{z}_3)=-1$, $\mathrm{H}(\eta \beta-\eta \beta_m)\in [-1,0]$ and
$\dot{\beta}=L-[-1,0]L/\eta$. This means that $\beta$ increases to approach $|\hat{z}_3|$, leading to $\beta>\beta_m$. Therefore, $\beta=|\hat z_3|$ is kept for $\eta \beta <|\hat z_3|$.
Therefore, the positive invariance of the set $[\beta_m, +\infty)$
is proved for $\beta(t)$ governed by \eq{law}.

%
\newcounter{mytempeqncnt}
\begin{figure*}[!t]
\center
\normalsize
\setcounter{mytempeqncnt}{\value{equation}}
\setcounter{equation}{49}
\begin{equation} \dfeq{Numeration}
\beta_i\!=\!\begin{cases}
\beta_{i-1}+\dfrac{1+\eta}{\eta}TL, &\mbox{if } \eta \beta_{i-1}\!<\!|\hat z_{3,i}|-(1+\eta) TL,\\ \dfrac{|\hat z_{3,i}|}{\eta}, &\mbox{if } |\hat z_{3,i}|-(1+\eta) TL \leq \eta \beta_{i-1} <|\hat z_{3,i}|-(1-\eta)TL,  \\ \beta_{i-1}+\dfrac{1-\eta}{\eta}TL, &\mbox{if } |\hat z_{3,i}|-(1-\eta)TL\leq \eta \beta_{i-1}<\eta \beta_m -(1-\eta) TL \\
\beta_m, &\mbox{if } \eta \beta_m -(1-\eta) TL \leq \eta \beta_{i-1} <\eta \beta_m +\eta TL \\
\beta_{i-1}-TL  &\mbox{if } \eta \beta_{i-1} \geq\eta \beta_m +\eta TL
\end{cases}
\end{equation}
\setcounter{equation}{\value{mytempeqncnt}}
\hrulefill
\end{figure*}

The rest of the claims can be proved by employing the argument
separating the dynamics into three phases.
In the first phase, the sliding $e=0$, i.e., $\hat z_3=\rho_0(z,t)$ is achieved regardless of $z\neq 0$. Notice that
property \eq{Gains} gives $L>L_2\geq|\rho_1(z,t)|$, which, in turn,
implies $e=0$ is achieved in finite time \cite{Levant_2003_higher}. Let $t_e$ be a real number defined as the finite time of convergence in the first phase.
The second phase of achieving the sliding mode $\delta:=\eta \beta-|\hat z_3|=0$ can be verified for the time interval $[t_e,\infty)$ as follows.

%

%
Let $V(\delta)=\delta^2/2$. From the above explanations and analysis, $\beta\in [\beta_m,+\infty ]$ is a invariant set for almost all the time. Then one has $\dot \beta =-L\sgn(\delta)$ and its
derivative yields
\begin{align}
\dot V&=\delta \dot \delta=\delta(-\eta L\sgn(\delta)-|\dot {\hat z}_3|) \nonumber\\
&=-|\delta|(\eta L + |\dot{\hat z}_3|\sgn(\delta)) \nonumber\\
&\leq  -|\delta|(\eta L -L_2)\nonumber \\
&<-\sqrt{2V(\delta)}(\eta L-L_2)<0.
\dfeq{Vdelta1ag}
\end{align}
Therefore, $\delta=0$ is
for $t\in[t_\delta,\infty)$, where
\begin{align}
t_\delta=\frac{\sqrt{2V(\delta(0))}}{\eta L-L_2} \in [t_e,+\infty).
\end{align}
 For $t\in[t_\delta,\infty)$, one has $\delta=0$, i.e., $\eta \beta=|\hat z_3|=|\rho_0(z,t)|$ for $\beta>\beta_m$. Combining with the previous analysis, one can conclude that for $t\in [t_\delta, +\infty]$, \eq{Utkin_kTW} is achieved, which is independent from the state $z=0$.

In the third phase, from \eq{Utkin_kTW} and Theorem~\ref{thm1}, one can conclude
 that there exist a $t_z\in [t_\delta, +\infty)$ such that $z=0$ is achieved in finite time $t\in [t_z,+\infty)$.
\qed
\end{pf}

\begin{rem}
Forward Euler integrations of \eq{law} can cause chattering on $\beta$, specially, for large value of $L$ and large time step sizes.
With implicit Euler discretization, one can achieve
a chattering free integration of $\beta$ \cite{Huber_2014_Twisting}. For simplicity, here only the first two terms in \eq{lawc} is considered. The implicit Euler methods for the three terms in \eq{lawc} can be attained by similar methods shown here.
By multiplying $\eta$ on both sides of \eq{lawc} and discretization \eq{lawc} with the implicit Euler method:
\begin{align} \dfeq{Discrete_form}
\eta \beta_i \in\eta\beta_{i-1}-\eta TL\mathrm{ {sgn}}(\eta \beta_i-|\hat z_{3,i}|)-
TL\mathrm{ H} (\eta \beta_i-\eta \beta_m) \nonumber\\
\end{align}
where $T>0$ is the time step size, $\beta_i\!:=\!\beta(t_i)$, $\hat z_{3,i}\!:=\!\hat z_{3}(t_i)$, and $t_i\!:=\!t_0+iT,\forall i\in \bbN_+, t_0\in \bbR_+$. One should note that $\hat z_{3,i}$ is a known scalar and can be explicitly obtained by integrating \eq{Observer}.
The formulas like \eq{Discrete_form} is a form of multiple cascaded set-valued inclusions such as $\sgn(\cdot)$ and $\mathrm{H}(\cdot)$ defined as \eq{sgn} and \eq{H}, respectively, \cite{Xiong_2014_BK}. Such structure of differential inclusions always have solutions \cite{Huber_2014_Twisting}. It can be also transformed into many other standard forms like mixed linear complementarity problem (MLCP) \cite{Xiong_2014_BK} and affine variational inequality (AVI) \cite{Huber_2014_Twisting}, which can be solved by many well-researched algorithms, solvers and even enumeration methods. Here, a geometry based enumeration method is illustrated as Fig.~\ref{enumeration}. The solution of $\beta_i$ is the vertical value of the cross point of two functions $y=\eta \beta_i-\eta \beta_{i-1}$ and $y \in -\eta TL\mathrm{\sgn}(\eta \beta_i-\hat z_{3,i})-TL\mathrm{ H}(\eta \beta_i-\eta \beta_m)$, depending on the known value of $\beta_{i-1}$. Here,only the solution for the case $|\hat z_{3,i}|\leq \eta \beta_m\leq +\infty  $ is given as \eq{Numeration}.
For the case of $\eta \beta_m< |\hat z_{3,i}|$, the solution can be obtained by exchanging $\eta \beta_m$ with $|\hat z_{3,i}|$ and $(1-\eta)TL$ with $\eta TL$, respectively.
\end{rem}

The idea of $\alpha(t)$ in \eq{law} can be seen by comparing
with \eq{gain_moreno}.
The gain $\alpha(t)$ is updated in accordance with the
magnitude of $\beta(t)$ needed to drive $z$ to the origin. Due to the predefined minimum value of $\beta_m$,
correspondingly, there is a minimum value $\alpha_m$ according to \eq{law}. The minimum values are very applicable to real situations. The differentiability of the perturbation $\rho_0(z,t)$ in \eq{Rho} is not always satisfied in practical applications, but it can be always divided into a differentiable partition and a non-differentiable partition. The minimum values of $\beta_m$ and $\alpha_m$ are required to deal with the non-differentiable partition, guaranteeing the robustness during the adaptation of STA. While the updated law \eq{law} is used for the differentiable partition. When the differentiable partition is small, the magnitudes of $\beta(t)$ and $\alpha(t)$ are also decreased
for preventing unnecessary chattering.
When large perturbations require large $\beta(t)$ and $\alpha(t)$, they are increased for enhancing the robustness.




\setcounter{equation}{50}

\begin{figure*}[t!]
\includegraphics[width=\textwidth]{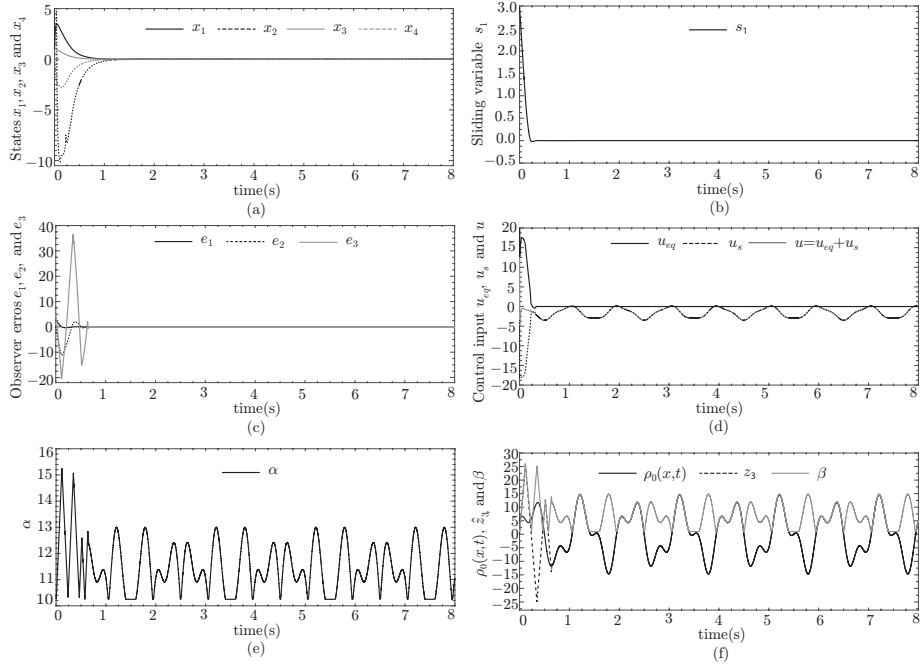}
  \caption{Adaptive equivalent control based sliding mode control (ECB-SMC) for the perturbed LTI system \eq{example} by applying the adaptive-gain law \eq{law} to the super-twisting algorithms (STA) \eq{STA}. Only the first line of gain updated law \eq{law} is integrated by the implicit Euler method \eq{Numeration} while \eq{example}, \eq{u_s}, and \eq{estimator} are integrated by the conventional forward Euler method with same time step size $T=0.0001\mathrm{s}$ (The time step size $T=0.0001\mathrm{s}$ is set to avoid the stiff problem in simulations due to the great value differences of matrix elements in \eq{example}). The derivative of the differentiable perturbation $\varphi$ in \eq{example} is selected as $GD\dot\varphi =\rho_0(x,t)=10\sin(2\pi t)+5\cos(5\pi t)$. The parameters are chosen as $\beta_m=1$, $\eta=0.99$, $L=200$, $h=1.01$, and $p=0.01$. The initial state of $x$ is $x(0)=[1,1,1,1]^T$.   }
  \label{fig_2}
\end{figure*}

\section{Example}
\label{sec:exam}
Consider
an industrial electromechanical emulator provided by Educational
Control Products (ECP), representing the important classes of
systems such as conveyors, machine tools, spindle drives, and
automated assembly machines \cite{Shyam_2015_Continuous}. It is consisted of a drive disk and
a payload disk and modeled as follows:
\begin{align}\dfeq{example}
\begin{bmatrix}
    \dot x_{1}        \\
    \dot x_{2}        \\
    \dot x_3         \\
    \dot x_{4}
\end{bmatrix}
&=\underbrace{
\begin{bmatrix}
    0 & 1 & 0 & 0 \\
    -209.6 &-2 & 838.4 & 1.7 \\
    0 & 0 & 0 & 1 \\
    77.9 & 0.15 & -311.8 & -2.47
\end{bmatrix}}_{A}
\begin{bmatrix}
     x_{1}        \\
     x_{2}        \\
      x_3         \\
     x_{4}
\end{bmatrix}+ \nonumber \\
&\quad\quad \quad\quad \quad\quad \underbrace{\begin{bmatrix}
    0       \\
    2306        \\
    0        \\
    0
\end{bmatrix}}_{B}u+D\varphi
\end{align}
where $x:=[x_1,x_2,x_3,x_4]^T$ is defied as the state, $x_1$ and $x_3$ are the angular position of the drive disk and
load disk, respectively, and $x_2$ and $x_4$ are the angular velocity
of the drive disk and load disk, respectively. For simplicity, it is assumed that $x$ is fully measurable and this assumption is reasonable because $x_1$ and $x_3$ can be obtained by using encoders while $x_2$ and $x_4$ can be estimated by designing two separated velocity observers \cite{Davila_2005_Oberver}. The control input $u$ is to drive $x:=[x_1,x_2,x_3,x_4]^T$ to zero in the presence of twice differentiable perturbation $\varphi \in \bbR$ with a constant perturbation matrix $D\in \bbR^{4\times1}$.

\par

Here, a simple sliding surface
\begin{align}\dfeq{surface}
s&=\underbrace{\begin{bmatrix}
     1 & 1/2306  & 1 &  1
\end{bmatrix}}_{G} x
\end{align}
is designed and its derivative is
\begin{align}\dfeq{surface_dot}
\dot s&=G\dot x= G[Ax+Bu+D\varphi ].
\end{align}
Similar to the conventional equivalent control based sliding mode control(ECB-SMC) \cite{Acary_2010_Euler}, the control input $u$ is divided into the continuous control $u_{c}$ and sliding mode control $u_s$, i.e., $u=u_{c}+u_s$. Here, the proposed adaptive STA is employed as the sliding mode control instead of the first order sliding mode control. Comparing to the conventional first oder sliding mode control, the adaptive STA can achieve a second order accuracy, i.e., $s=\dot s=0$, with the chattering being greatly attenuated due to the absolute continuity of STA and adaptively attenuated gains. The equivalent control and sliding mode control are
\bsbeq
\begin{align}\dfeq{u_eq}
&u_{c}=-(GB)^{-1}GAx \\
&u_s=(GB)^{-1}[-\alpha (t) |s|^{1/2}\sgn(s)+\sigma] \dfeq{u_s}\\
&\dot \sigma\in -\beta(t) \sgn(s).
\end{align}
\esbeq
By substituting $u$ in \eq{surface_dot} with \eq{u_eq} and \eq{u_s}, defining $s:=s_1$, $s_2:=\sigma+GD\varphi$ and assuming that the perturbation $|\rho_0(x,t)|:=|GD\dot \varphi |\leq L_1$, $|\rho_1(x,t)|:=|GD\ddot \varphi| \leq L_2$ for all $x\in\bbR^4$ and $t\in\bbR_+$, one can obtain
\bsbeq \dfeq{sta_exam}
\begin{align}
\dot s_1&= s_2-\alpha(t) |s_1|^{1/2}\sgn(s_1)\\
\dot s_2 &\in-\beta(t) \sgn(s_1)+ \rho_0(x,t)
\end{align}
\esbeq
in the form of \eq{STA}. According to the perturbation observer \eq{Observer}, the perturbation $\rho_0(x,t)$ can be estimated by employing the following dynamics:
\bsbeq \dfeq{estimator}
\begin{align}
\dot {\hat z}_1&=\hat z_2-\alpha(t) |s_1|^{1/2}\sgn(s_1)+k_1|e_1|^{2/3}\sgn(e_1)\\
\dot {\hat z}_2&=-\beta(t) \sgn(s_1)+k_2|e_1|^{1/3}\sgn(e_1)+\hat z_3 \\
\dot {\hat z}_3&\in k_3\sgn(e_1)
\end{align}
\esbeq
where $e_1\!:=\!s_1-\hat z_1$. By selecting the gains $k_1,k_2$ and $k_3$ as in \eq{gains}, and defining $e_2\!:=\!s_2-\hat z_2$ and $e_3\!:=\!\rho_0(x,t)-\hat z_3$, one has the error dynamics of perturbation observer:
\bsbeq \dfeq{estimator_error}
\begin{align}
\dot e_1&=-3L^{1/3}|e_1|^{2/3}\sgn(e_1)+e_2\\
\dot e_2&=-1.5\sqrt{3}L^{2/3}|e_1|^{1/3}\sgn(e_1)+e_3 \\
\dot e_3&\in -1.1L\sgn(e_1)+\rho_1(x,t)
\end{align}
\esbeq
in the form of \eq{Error_standard}. For large enough value of $L>L_2\geq|\rho_1(x,t)|$, the error $e:=[e_1,e_2,e_3]^T$ will disappear in finite time. Then one has $\hat z_3=\rho_0(x,t)$, of which the magnitude is used as the tracking target of $\beta$ in \eq{sta_exam} by the update law \eq{law}.

\par

One can obtain as small as possible control input $u$ by integrating the controller \eq{u_eq} and \eq{u_s} with the conventional forward Euler method and by integrating \eq{law} with the backward Euler method \eq{Numeration}. The gains $\alpha$ and $\beta$ will be self-adjusted by the law \eq{law} to as small as possible but large enough to counteract the perturbation $\rho_0(x,t)$. The chattering caused by the very large value of $L$ in \eq{law} is removed by the geometry based backward Euler integration method \eq{Numeration}.

\par

Fig.~\ref{fig_2} shows the simulation results of
designing an equivalent control based sliding mode control (ECB-SMC) with the proposed adaptive super-twisting algorithm (STA) \eq{STA} for the perturbed LTI system \eq{example}. Fig.~\ref{fig_2}(a) shows that the state $x$ asymptotically converges to zero while Fig.~\ref{fig_2}(b) shows that the sliding variable $s_1$ converges to zero in finite time in the presence of perturbation $\varphi$. The sliding variable $s_1$ is a polynomial composition of $x_i, i\in \{1,2,3,4\}$, which is a very common design for a higher order system like \eq{example} controlled by a relatively lower sliding mode control such as STA \eq{STA}.
Fig.~\ref{fig_2}(c) shows that in a finite time, the observer error \eq{estimator_error} converges to zero, which is earlier than the convergence of the state $x$ (less than $1$ second). This implies that the adaption of the gains $\alpha$ and $\beta$ is independent from the convergence of the STA \eq{estimator}. This property is different from conventional adaption laws based on low-pass filters, which begin to adjust the gains only after the convergence of STA.
\par

Due to $s_1=0$ before the convergence of $e$ in \eq{estimator_error} to zero, the small magnitude $\eta \beta=|\rho_0(x,t)|$ or $\beta=\beta_m$, does not affect the convergence time $t_z$ for $s_1=0$. This phenomena can be observed by comparing Fig.~\ref{fig_2}(b),(d), and (f). This is important property because the quicker convergence of $\eta \beta=|\rho_0(x,t)|$ or $\beta=\beta_m$, which is a small magnitude to counteract the perturbation $\varphi$, can result in a long convergence of $s_1=0$. The reason is that with the same initial state $s(0)$, the convergence of \eq{sta_exam} is inverse proportional to $\beta$. To achieve a quicker convergence for $s=0$ in \eq{sta_exam}, one can increase $L$ and decrease $\eta$, leading to a larger gain $\beta=|\rho_0(x,t)|/\eta$.

\par
In Fig.~\ref{fig_2}(e)(f), it is observed that
the adaptation \eq{law} adjusts $\beta(t)$ to the level of perturbation $\rho_0(x,t)$ and changes $\alpha(t)$ proportionally to $\beta(t)$.
The two gains $\alpha(t)$ and $\beta(t)$ adjusted by
\eq{law} reduce as the magnitude of $\rho_0(x,t)$ decreases.
When $\rho_0(x,t)$ is small, i.e., $|\rho_0(x,t)|< \eta \beta_m$ is satisfied
in Theorem \ref{thm2},
$\beta(t)$ converges to $\beta_m$. Staying at the predefined minimum value to guarantee the robustnesses. When $|\rho_0(x,t)|>\eta \beta_m$, a magnitude $\beta=|\rho_0(x,t)|/\eta$, as small as possible but large enough to grantee the robustness, is achieved.

\section{Conclusion}
\label{sec:con}
In this paper, an adaption methodology of super-twisting
algorithm is developed based on third-order observer and
Lyapunov function approaches. It can realize the possibly
minimum level of gains while keeping the robustness of sliding
mode control without using low-pass filters to estimate the boundary of perturbations. The
efficacy of the proposed adaptive super-twisting algorithm is
confirmed by designing an equivalent control based sliding mode control (ECB-SMC) with the proposed adaptive super-twisting algorithm for a benchmark system in the literature.

\par

The same mechanism can be applied to adapt the gains of higher order sliding mode (HOSM). Further study may focus on various applications of adaptive HOSM, such as adaptive differentiators, adaptive sliding mode controllers, adaptive observers, and adaptive filters.

\section{Acknowledgment}

This work was supported by the Natural Science Foundation of China (Grant No. 11702073), Shenzhen Key Lab Fund of Mechanisms and Control in Aerospace (Grant No. ZDSYS201703031002066), and by the Basic Research Plan of Shenzhen (JCYJ20170413112645981).







\bibliography{mybibfile}

\end{document}